\documentclass[conference]{IEEEtran}
\IEEEoverridecommandlockouts
\usepackage[bookmarks=false]{hyperref}
\usepackage{cite}
\usepackage{amsmath,amssymb,amsfonts}
\usepackage{algorithmic}
\usepackage{graphicx}
\usepackage{textcomp}
\usepackage{xcolor}
\usepackage{orcidlink}
\usepackage{subfig}
\usepackage{multirow}
\def\BibTeX{{\rm B\kern-.05em{\sc i\kern-.025em b}\kern-.08em
    T\kern-.1667em\lower.7ex\hbox{E}\kern-.125emX}}
\begin{document}

\title{\fontsize{20}{22}\selectfont \textbf{FERMI-ML}: A \textbf{F}lexible and \textbf{R}esource-\textbf{E}fficient \textbf{M}emory-\textbf{I}n-Situ SRAM Macro for TinyML acceleration

\thanks{\textsuperscript{\textdagger}Both authors contributed equally to this work.}
\thanks{This work was supported by the Special Manpower Development Program for Chip to Start-Up (SMDP-C2S), Ministry of Electronics and Information Technology (MeitY), Government of India, under Grant EE-9/2/21 - R\&D-E.}
}

\author{
    \IEEEauthorblockN{
        Mukul Lokhande\textsuperscript{\textdagger, \IEEEauthorrefmark{1},\orcidlink{0009-0001-8903-5159}}, Member IEEE, 
        Akash Sankhe\textsuperscript{\textdagger, \IEEEauthorrefmark{1}, \orcidlink{0009-0006-2397-5670}},
        S. V. Jaya Chand\textsuperscript{\IEEEauthorrefmark{1}, \orcidlink{0009-0007-6855-7859}},\\
        and Santosh Kumar Vishvakarma\textsuperscript{\IEEEauthorrefmark{1}, \orcidlink{0000-0003-4223-0077}}, Senior Member IEEE. 
    }
    \IEEEauthorblockA{\IEEEauthorrefmark{1}Indian Institute of Technology Indore-453552, India}
    Email: skvishvakarma@iiti.ac.in \textbf{(Corresponding Author)}
}

\maketitle

\begin{abstract}
The growing demand for low-power and area-efficient TinyML inference on AIoT devices necessitates memory architectures that minimise data movement while sustaining high computational efficiency. This paper presents FERMI-ML, a Flexible and Resource-Efficient Memory-In-Situ (MIS) SRAM macro designed for TinyML acceleration. The proposed 9T XNOR-based RX9T bit-cell integrates a 5T storage cell with a 4T XNOR compute unit, enabling variable-precision MAC and CAM operations within the same array. A 22-transistor (C22T) compressor-tree-based accumulator facilitates logarithmic 1–64-bit MAC computation with reduced delay and power compared to conventional adder trees. The 4 KB macro achieves dual functionality for in-situ computation and CAM-based lookup operations, supporting Posit-4/FP-4 precision. Post-layout results at 65 nm show operation at 350 MHz with 0.9 V, delivering a throughput of 1.93 TOPS and an energy efficiency of 364 TOPS/W, while maintaining a Quality-of-Result (QoR) above 97.5\% with Inception-V4 and ResNet-18. FERMI-ML thus demonstrates a compact, reconfigurable, and energy-aware digital Memory-In-Situ macro capable of supporting mixed-precision TinyML workloads.
\end{abstract}


\section{Introduction}
Machine learning (ML) applications have found use cases in every AIoT device, for incorporating human-like features such as face recognition, specialised recommendations, and voice commands. Deep neural networks lie at the centre of these systems, assisting them in identifying complex, hidden patterns based on sensor data and facilitating human-machine coexistence. However, the resource-efficient implementation of these systems has remained the fundamental challenge over the years. The workload characterisation reveals the dominance of multiply-accumulate (MAC) operations, such as VGG-16 and ViT-G, which require 15.5 billion and 2.86 GMACs, respectively [1],[2]. Furthermore, it should be noted that the data transfer between the computational unit and the storage limits the performance, a phenomenon known as the von Neumann bottleneck. For example, VGG-16 and ViT-G have 138 million weights and 184 billion parameters for 224x224x3 image classification and object detection inference on ImageNet. 

Recent industry efforts [3]-[8] have converged with Memory-In-situ for TinyML acceleration as a promising solution for resource-efficient matrix multiplication operations. The key focus has been to eliminate the energy-hungry data transfer and incorporate the computational logic (AND/NOR/XNOR), adjacent tothe  memory-bitcell [9]. Prior analog PIM approaches, performance remains impacted by non-linear PVT variations, signal margin degradation, resource-overhead associated with ADCs/adder tree, and macro-applicability limited to 4-bit precision for advanced AI models. However, the recent digital PIM macros have improved robustness, variable-bit precision and satisfiable accuracy-hardware resource trade-offs. Thus, we aim for the flexible digital Memory-In-Situ SRAM macro solution. 

Prior works primarily incorporated AND, NOR or XNOR logic with 6T/8T storage SRAM. However, the 6T AND logic [10]-[13], 4T NOR logic and 2T for inversion [14],[15] and 4T XNOR logic [16]-[20] reduce the storage density significantly, in comparison with same area memory bank. Furthermore, adder-trees contribute up to 79\% power and 45\% area overhead [15], making digital solutions inconvenient to replace as SRAM caches. Prior compressors [12],[16],[21]-[26] focused on resource-efficiency of an accumulation tree based on an accuracy-resources trade-off. The key observation was, The storage density reduced up-to 240 KB in place 1MB L3-cache. Thus, the primary motivation of this work is to enhance storage and compute density, reduce the power consumption associated with accumulation, and explore the opportunity for flexibility within the SRAM-In-situ macro. 

\begin{figure*}[!t]
    \centering
    \includegraphics[width=0.85\textwidth]{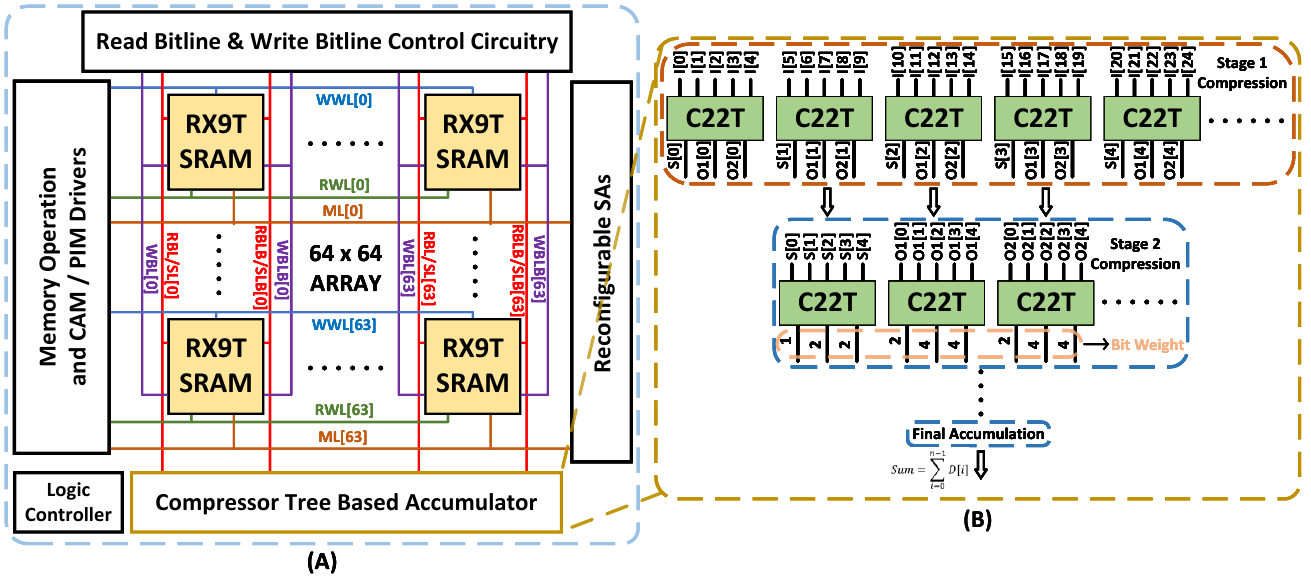}
    \caption{Detailed circuitry showing (a) Proposed Memory-In-situ SRAM bank architecture, with detailed (b) novel resource-efficient compressor tree Structure}
    \label{fig:macro}
\end{figure*}

The key contributions of this work are:
\begin{enumerate}
    \item \textbf{Resource-efficient 9T XNOR-based (RX9T) Variable-precision MAC compute:} This work utilised ultra-low power 5T storage bit-cell and 4T XNOR circuitry for XAC/MAC operation. The 22T compressor (C22T) tree-based accumulator supports logarithmic 1/2/4/8/16/32/64-bit precision-based MAC computation, which is latency-effective compared to prior adder-tree accumulation. 

    \item \textbf{Read-decoupled row-wise content-addressable memory (CAM) operation:} The approach allows utilising the same 4T XNOR for CAM operation, matrix transposition, and follows a state-of-the-art look-up-table-based approach for non-linear activation functions. This also allows configurability for Posit-4/FP-4 precision within the same LUT structure. 

    \item \textbf{Flexible Memory-In-Situ 4Kb SRAM macro for TinyML acceleration:} The proposed flexible macro operates at 350 MHz@0.9V, benefits from the RX9T and C22T, with energy for PIM of 17.65 fJ/bit and CAM of 0.55 fJ/search/bit and translates to post-layout performance of enhanced energy-efficiency 364 TOPS/W and compute density 4.58 TOPS/mm\textsuperscript{2}.

\end{enumerate}

\section{Proposed approach}
AI inference is computed with the proposed Memory-In-situ SRAM macro, with detailed circuitry for a bank shown in Fig. \ref{fig:macro}. The fundamental elements consist of an SRAM array with RX9T, a 4:2 C-22 T-based accumulator, a precision-reconfigurable sense amplifier required for conventional memory and CAM operations, a Bit-line control circuitry, a PIM driver circuitry and bank control logic. For simplicity, we have demonstrated a 64x64 array with a total size of 4 KB. The architecture discussed is scalable and can be articulated as needed. The key elements include novel RX9T, C-22T, which are discussed in the upcoming subsections. 

\subsection{Resource-efficient 9T XNOR SRAM bit-cell}

The resource efficiency of SRAM-bitcell is dependent on storage transistors (cross-coupled inverters), access mechanisms, and compute logic. Prior works utilised heavy logic of up to 6T for AND [10], [11], 6T for NOR [14], [15], and 4T for XNOR [16], [19] with a complicated access mechanism. Even prior works focused on multiple functionalities, such as AND/XNOR [22] and AND/NOR [27]; however, they fail to address PIM and CAM operations with a single reconfigurable logic. Alternatively, ReRAM-based individual works with XNOR, focusing on PIM [28] and CAM [29]; however, it also lacks a unified approach. 

\begin{figure}[!h]
    \centering
    \includegraphics[width=0.825\columnwidth]{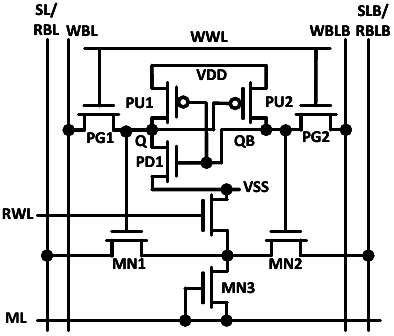}
    \caption{Schematic for Resource-efficient 9T XNOR SRAM bit-cell for Memory-In-Situ processing.}
    \label{fig:bitcell}
\end{figure}

Different SRAM-bitcells with focus on ultra-low power operation (8T/9T) [30], [31], resource-efficiency (A5T) [32] were proposed synonymous to conventional 6T and read-decoupled 8T (RD8T) storage SRAM. We utilised A5T with dual-port operation and low-power consumption compared to RD8T to replace the conventional 8T [10], [17] for storage. We utilised 4T for XNOR with decoupled read-write for compute and match line for CAM operation, and formulated the proposed 9T XNOR bit-cell shown in Fig. 2. The detailed logical operation is briefed in Table I.

\subsubsection{Normal Memory Mode}
Table I illustrates the write operation of the proposed cell, which is nearly identical to the traditional 8T SRAM operation. When the write word line (WWL) is asserted, data from the write bitline (WBL) and the write bitline bar (WBLB) are stored in the bitcell. The bitcell features row-wise and column-wise read modes. For the rowwise read, the read SA is configured as a differential SA. All MLs across the array are driven to low voltage, which disables the MN3 transistor. The read wordline (RWL) of the selected row is asserted, and the RWLs of unselected rows are de-asserted. Depending on the data stored, the read bitline (RBL) or read bitline bar (RBLB) discharges. The differential SAs below the array compare the voltages of RBLs and RBLBs to read the stored data from the selected row. For column-wise read operation, all the RWLs in the array are driven to a low voltage level, turning off the MN4 transistor. Then the MLs are pre-charged to high voltage, and the RBL and RBLB of the selected columns are driven to high voltage and low voltage pulses, respectively. When the data stored is 1, the ML remains at high voltage, and when the data stored is 0, the ML discharges through the MN3-MN1-GND (RBL) path. The single-ended SAs on the right of the array compare the ML voltage with a reference voltage (100mV below VDD) to obtain the results. The benefits of this approach are two-fold. After the data is written row-wise, the transpose of the matrix can be easily obtained by reading the data column-wise, thus significantly enhancing efficiency. 

\begin{table}[!t]
\caption{The operation table for bit-cell showing SRAM (read, write and hold), multiplication and CAM operation in terms of logical values.}
\label{tab:operation}
\renewcommand{\arraystretch}{1.35}
\resizebox{\columnwidth}{!}{%
\begin{tabular}{|l|ll|l|l|l|l|}
\hline
\textbf{Operation} & \multicolumn{2}{c|}{\textbf{Storage (Read)}} & \multirow{2}{*}{\textbf{Storage (Write)}} & \multirow{2}{*}{\textbf{SRAM Hold}} & \multirow{2}{*}{\textbf{Multiplication}} & \multirow{2}{*}{\textbf{CAM}} \\ \cline{1-3}
\textbf{Control Line} & \multicolumn{1}{c|}{\textbf{Row}} & \textbf{Column} &  &  &  &  \\ \hline
\textbf{WWL} & \multicolumn{1}{c|}{Low (0)} & Low (0) & High (1) & Low (0) & Low (0) & Low (0) \\ \hline
\textbf{WBL} & \multicolumn{1}{c|}{NC} & NC & \begin{tabular}[c]{@{}c@{}}High (1) for Write (1)\\ Low (0) for Write (0)\end{tabular} & NC & NC & NC \\ \hline
\textbf{WBLB} & \multicolumn{1}{c|}{NC} & NC & \begin{tabular}[c]{@{}c@{}}Low (0) for Write (1)\\ High (1) for Write (0)\end{tabular} & NC & NC & NC \\ \hline
\textbf{SL/RBL} & \multicolumn{1}{c|}{High (1)} & High (1) & NC & NC & \begin{tabular}[c]{@{}c@{}}High (1) for Search (1)\\ Low (0) for Search (0)\end{tabular} & High (1) \\ \hline
\textbf{RBLB/SLB} & \multicolumn{1}{c|}{High (1)} & Low (0) & NC & NC & \begin{tabular}[c]{@{}c@{}}High (1) for Search (0)\\ Low (0) for Search (1)\end{tabular} & High (1) \\ \hline
\textbf{RWL} & \multicolumn{1}{c|}{High (1)} & Low (0) & Low (0) & Low (0) & Low (0) & High (1) \\ \hline
\textbf{ML/HBL} & \multicolumn{1}{c|}{Low (0)} & High (1) & NC & High (1) & High (1) & Low (0) \\ \hline
\end{tabular}}
\end{table}

\subsubsection{Content Addressable Memory Mode}
When configured to be used as a NOR content-addressable memory (CAM), the bitcell forms a discharge path for each mismatched data in the row. As the discharge time must be greater than the worst delay case (1-bit mismatch case), the ML with multi-bit mismatch discharges to a low level. To solve the issue of mismatch line swing, we directly connect the gate and source of the MN3 transistor to the ML, thus creating a self-feedback system to control the discharge process.

\textbf{Binary CAM (BCAM) Mode}
Like traditional CAMs, the search operation is conducted rowwise, which is compatible with the traditional write mode. Each 9T bitcell is a single BACM. The search data are compared with the data stored in each row of the array. If a data match is found, the ML of the matched row remains at high voltage. If a data mismatch is detected, the ML discharges, indicating a failed search operation. For BCAM mode, the RBL and RBLB are configured as SL and SLB, respectively, and all the read word-lines are connected to low voltages, thereby turning off the MN4. The ML are pre-charged to high voltage, and if the search criteria is 1, then SL and SLB are driven to high and low voltage, respectively, and vice versa. Thus, when the stored data is 1, then MN1 is on and MN2 is off, and there is no discharge path. When the stored data is 0, MN1 is off and MN2 is on, thus providing a discharge path from ML-MN2-SLB-GND and the SAs located to the right of the array. The SAs sense and compare the ML voltage with the reference voltage to produce a suitable output. Table I shows the BCAM operation of the proposed bitcell.

\textbf{Ternary CAM (TCAM) Mode}
TCAM adds an extra uncertain state X on top of the 0 and 1 states in BCAM. Thus, to represent 0,1 and X, we need two SRAM cells per TCAM. If both bitcell hold the value 0, then it represents state 0. If both bitcell stores 1, then it represents the state. If the first bitcell holds value 0 and the second bitcell holds value 1, then it represents an uncertain state X. But if the first bitcell stores 1 and the second stores 0, then the discharge will happen irrespective of the search condition. The RBL of the first bitcell is configured as SL, and the RBLB of the second bitcell is considered as SLB. The RBLB of the first bitcell and the RBL of the second bitcell are kept at high voltages to mitigate their impact on search results. Table I shows the TCAM operation of the proposed bitcell. When search data is 0, SL0 and SLB1 are driven to 0 and 1, respectively, and vice versa.

\begin{figure}[!t]
    \centering
    \includegraphics[width=0.875\columnwidth]{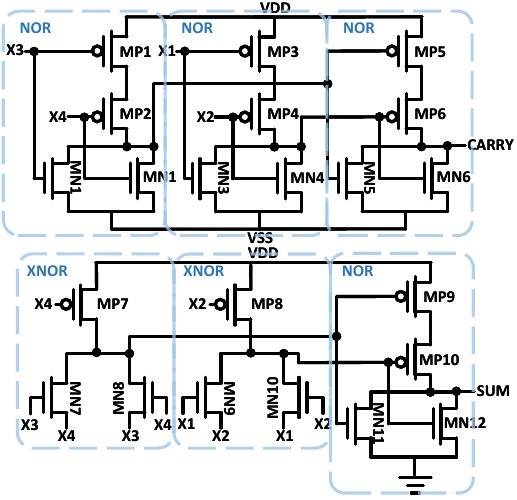}
    \caption{Schematic for novel 4:2 compressor with 22T, for faster and power-efficient accumulation.}
    \label{fig:comp-schematic}
\end{figure}

\subsubsection{Processing-In-Memory Mode}
For PIM mode, logic operations are implemented through simultaneous multiple row activation and require 2 single-ended SAs for a Boolean operation. Thus, we use the RSA located at the bottom of the array as two single-ended SAs. The reference voltage for the SAs is kept at 100mV below VDD. The logic operations performed by the bitcell are similar to those in [18] and [19]; the only difference is that the MN3 transistor does not participate in the discharge process, so it is turned off by discharging the ML line. Table I shows the PIM operation mode signals. For the unused rows, the read wordlines are connected to GND to prevent any impact on the output results. The readbitlines are precharged to supply voltage and the RWLs of the selected rows are set to high voltage. For example, if the stored data in both bitcell is 1, then the RBL will discharge through the MN1-MN4-GND path while the RBLB remains high, and if the stored data in both bitcell is 0, then vice versa happens. If the data stored is either 01 or 10 then both RBL and RBLB have a discharge path and the output is obtained from the difference between the voltage of bitlines and the reference voltage. Thus, by activating multiple rows simultaneously, we can perform multiple operations directly in memory.

\subsection{Fast and Power-Efficient 22T-based 4:2 Compressor}

Adder-tree-based accumulation was the key bottleneck that contributed up to 79\% of the total power in 4-bit macros and up to 82\% in 8-bit macros [15]. Subsequently, the RCA-based tree also affects the delay and area consumption. This was addressed in Flex-DCIM [10] using an 8-bit power-gated FA-26T, which resulted in a 12\% reduction in area and a 35\% reduction in power consumption. Our approach focused on logarithmic barrel-shifter and compressor-based accumulation, with a 22T 4:2 compressor for resource efficiency. This approach saves a significant transistor count compared to the prior 56T conventional compressor design and 40T [22] SoTA compressor. The compressor circuit is shown in Fig. 3, which follows PTL logic in stage 1 and CMOS logic in stage 2, effectively minimising the number of transistors. For fairer comparison, we have compared with prior compressor-structures [12], [21]-[23], [25] and full-adder equivalent (1 Approx-FA and 1 accurate FA), and found faster and more power-efficient with an error rate of 2.23\%, NMED 0.38 and MRED 0.256 for 8-bit accumulations and justified performance for image smoothing and edge detection. 

\begin{figure}[!t]
\centering
\subfloat[]{\includegraphics[width=0.95\columnwidth]{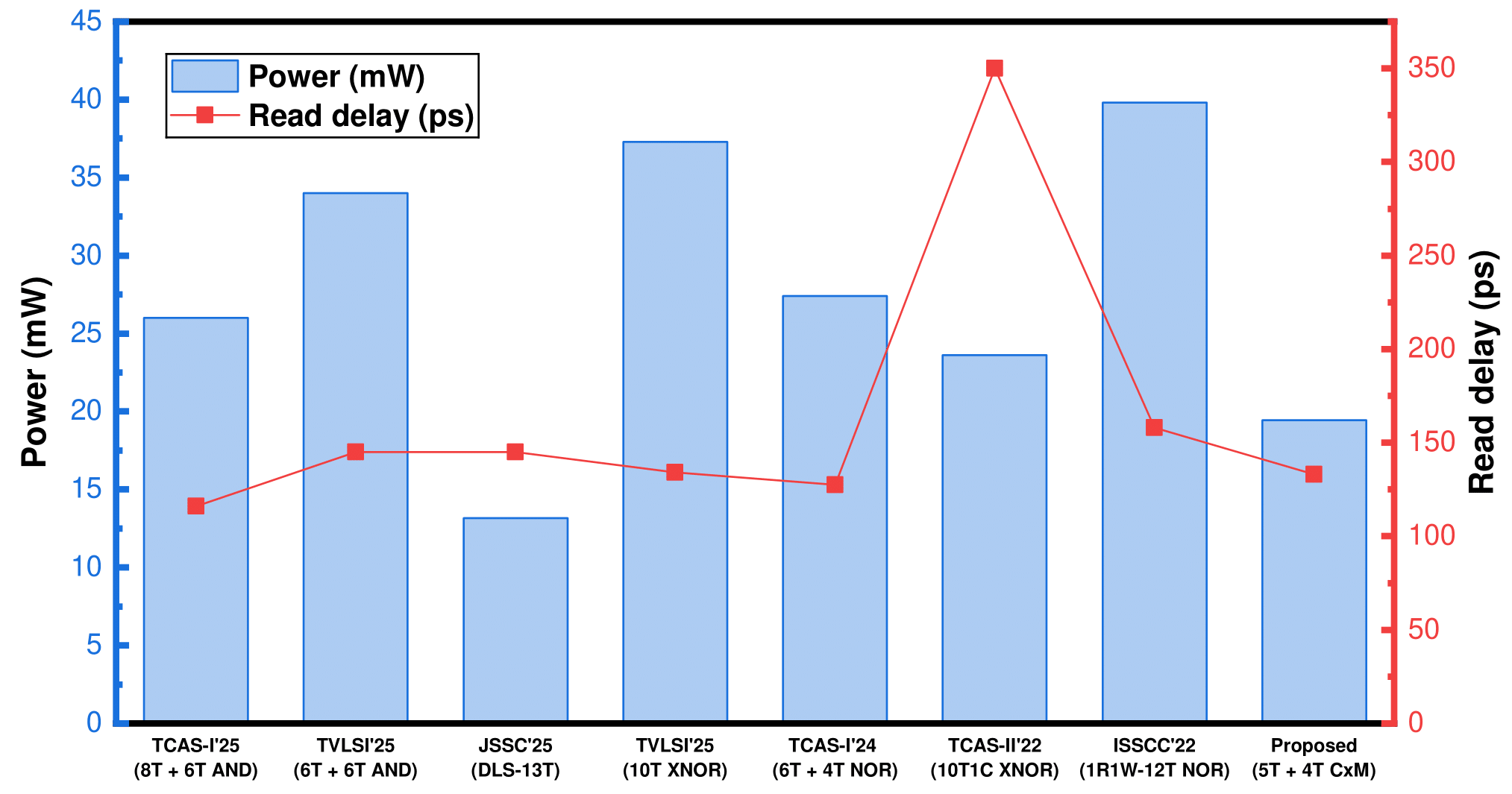}}
\hfill
\subfloat[]{\includegraphics[width=0.95\columnwidth]{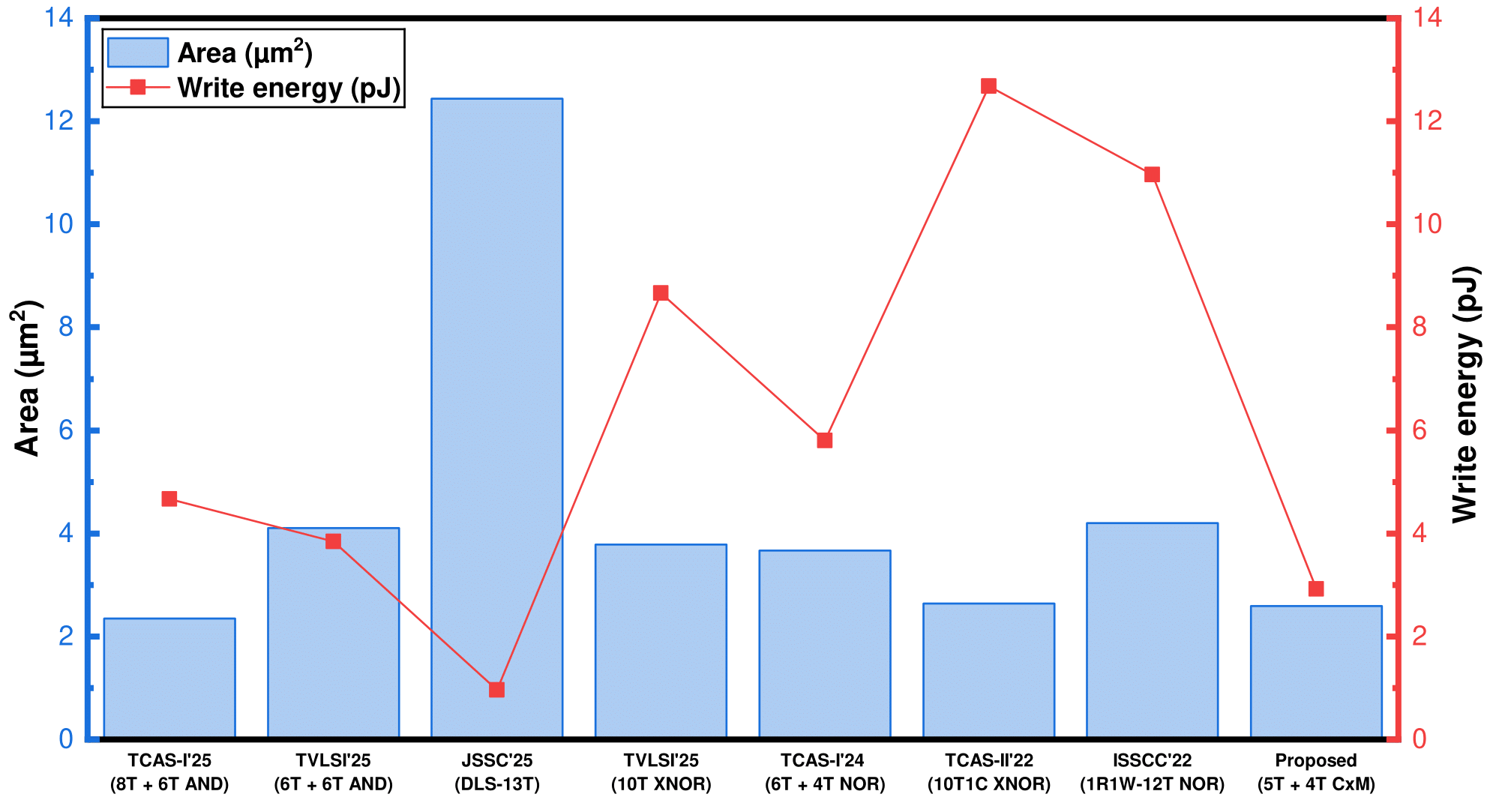}}
\caption{Post-layout performance comparison with State-of-the-Art Memory-In-Situ SRAM bitcells [10], [11], [14], [15], [17]-[20], [32], in terms of (a) Power and Read delay, (b) Area and Write energy.}
\end{figure}


\begin{figure}[!t]
    \centering
    \subfloat[]{\includegraphics[width=0.48\columnwidth, height=32.5mm]{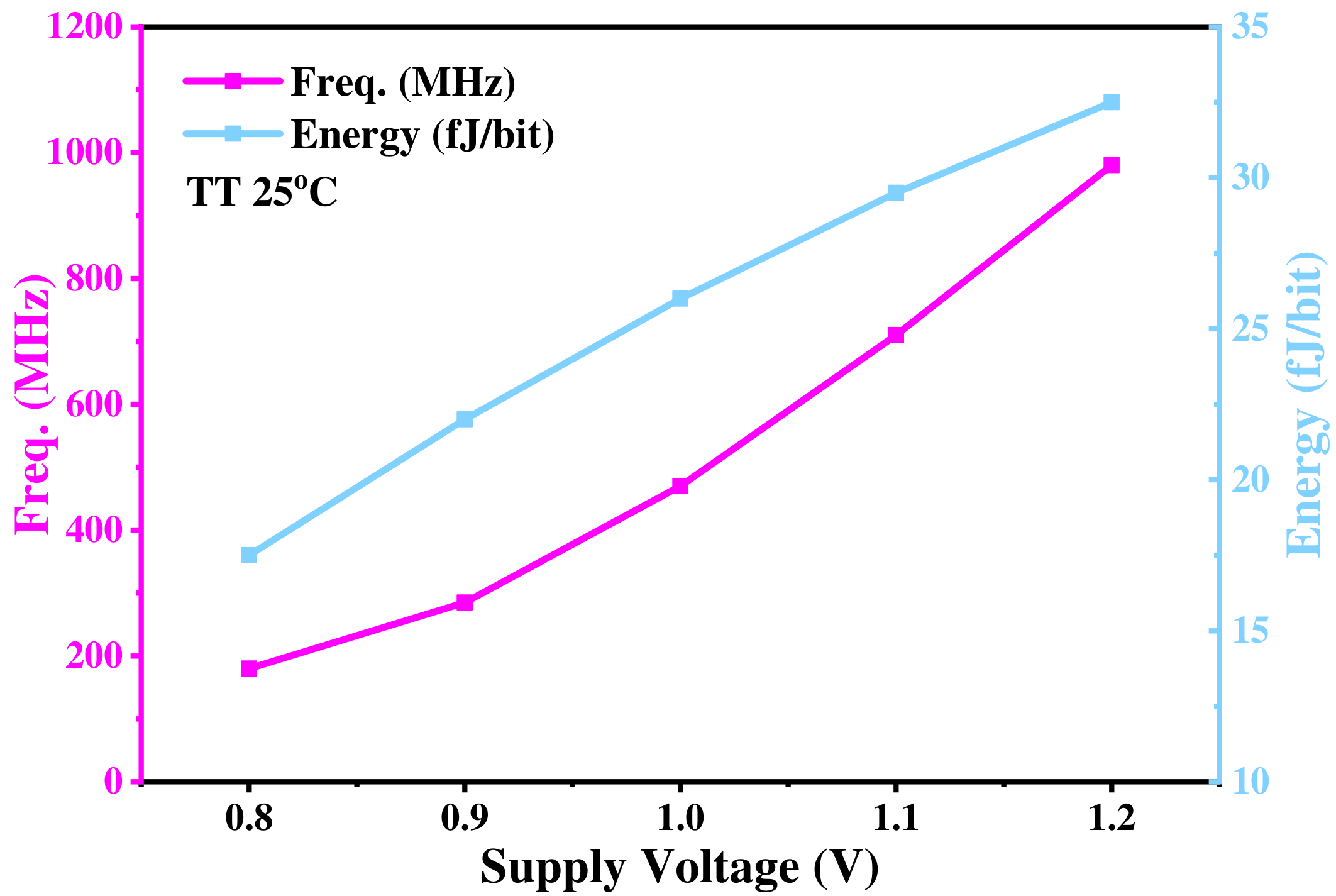}}
    \centering
    \subfloat[]{\includegraphics[width=0.48\columnwidth, height=32.5mm]{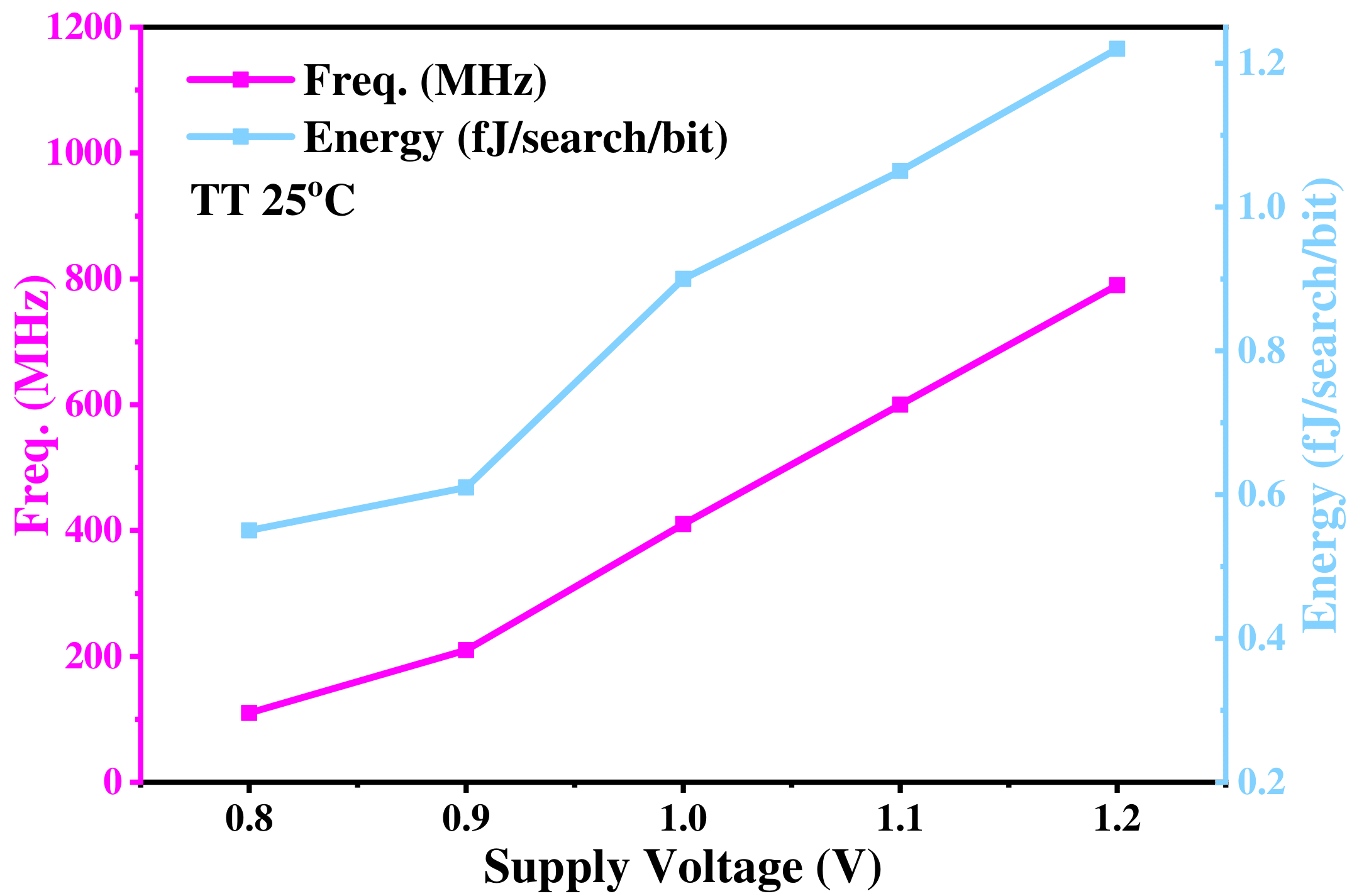}}
    \caption{Impact of Dynamic voltage-frequency scaling on energy-consumption of proposed SRAM macro for (a) Memory-In-situ XAC/MAC operation in matrix multiplication, (b) CAM operation for Look-up-table.}
\end{figure}

\begin{figure}[!t]
    \centering
    \includegraphics[width=0.925\columnwidth]{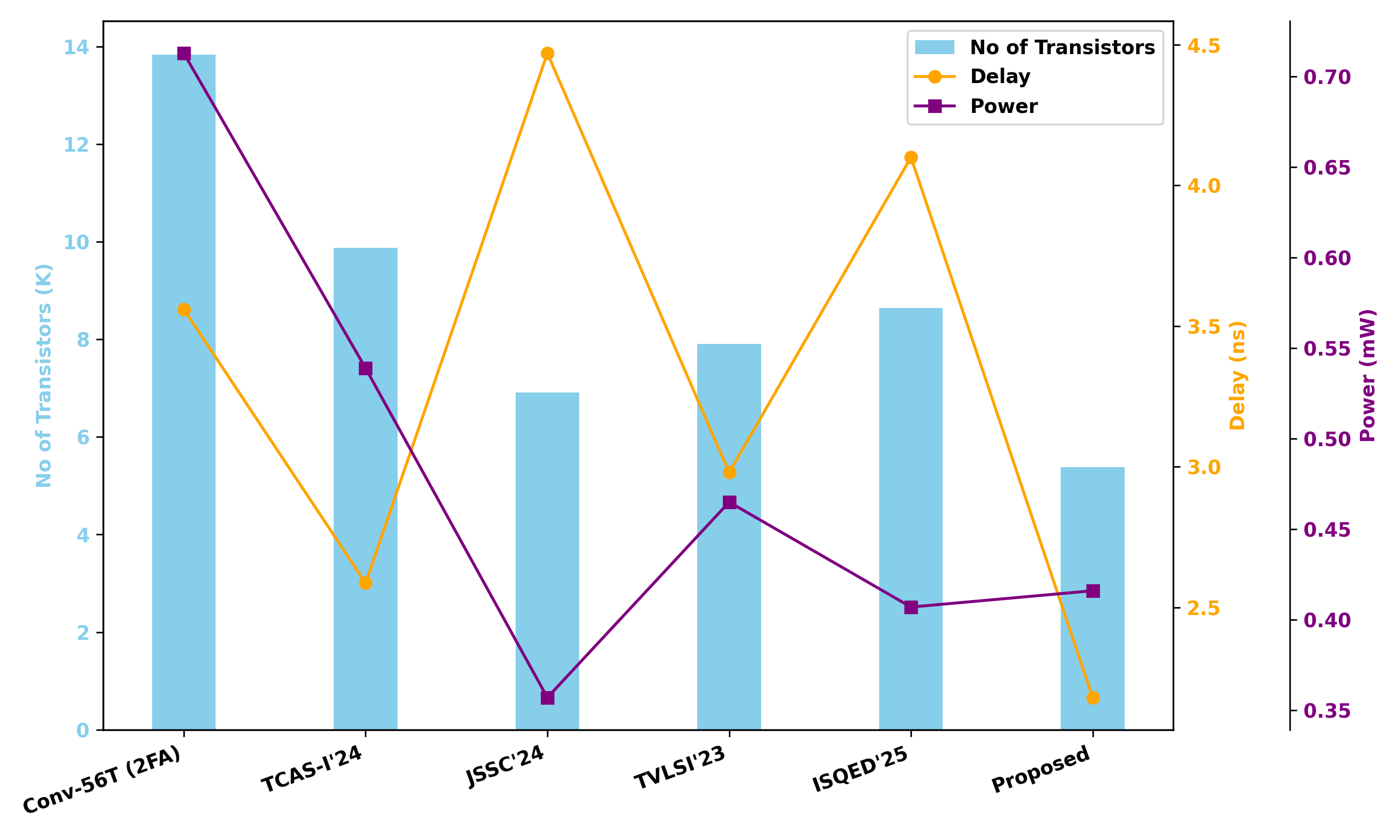}
    \caption{Performance comparison with different accumulation approaches [12], [13], [16], [21]-[25] in terms of Transistor count, Delay (ns) and Power ($\mu$W).}
    \label{fig:comp-perf}
\end{figure}

\section{Performance Evaluation}

The proposed 4 Kb macro is capable of variable-precision SIMD MAC compute, either 4096 1b/1b operations, 1024 2b/2b operations, 256 INT4 operations, 64 INT8 operations, 16 INT16 operations, or 4 INT32 operations in parallel, accompanied by compressor-based accumulation across multiple clock cycles. The scalability and re-programmability are handled in a sequential manner for tile-wide execution within a layer, followed by layer-wide precision for accuracy-hardware tradeoffs, based on control logic and PIM driver. The work utilises vector parallel ReLU activation and follows off-chip control circuitry similar to [13].

Different state-of-the-art Memory-in-situ SRAM bit cells are compared in Fig. 4 in terms of area, power, read delay, and write energy. The significance for the area is in translation to compute and storage density, while power is related to energy efficiency and throughput, which are impacted by read delay and write energy. Our approach shows a significantly comparable area with respect to prior approaches [10]-[15], [17]-[20] except for [33], which is 4.8 $\times$ larger compared to RX9T. This approach was found to be significantly power-efficient for a single multiplication cycle (from write data to multiplication output), where our approach saves power by a factor of 1.92 $\times$ and 1.22 $\times$ compared to prior XNOR approaches [17] and [20], respectively.

\begin{table*}[!t]
\caption{Edge AI PIM macro, compared in terms of scaled 16 Kb size for scalable TinyML acceleration.}
\label{tab:macro-comp}
\renewcommand{\arraystretch}{1.15}
\resizebox{\textwidth}{!}{%
\begin{tabular}{|l|c|c|c|c|c|c|c|}
\hline
 & \textbf{TCAS-I'25 [10]} & \textbf{ISQED'25 [13]} & \textbf{TCAS-I'24 [15]} & \textbf{TCAS-I'23 [34]} & \textbf{TNANO'23 [18]} & \multicolumn{1}{l|}{\textbf{DATE'25 [35]}} & \multicolumn{1}{l|}{\textbf{Proposed}} \\ \hline
\textbf{Tech. Node (nm)} & 65 & 65 & 55 & 65 & 65 & 65 & 65 \\ \hline
\textbf{Compute logic} & PIM & PIM & PIM & NMC & XAC & PIM & PIM/CAM \\ \hline
\textbf{Bit-cell size} & 8.75T & 8T & 8T & 7T & 10T & 8T & 9T \\ \hline
\textbf{Supply Voltage (V)} & 1 & 1.2 & 1.2 & 0.8-1.1 & 1.2 & 0.6-1.2 & 0.8-1.2 \\ \hline
\textbf{Macro Size (Kb)} & 64 & 16 & 64 & 80 & 16 & 4 & 4 \\ \hline
\textbf{Freq. (MHz)} & 400 & 250 & 200 & 200 & 25 & 40 & 350 \\ \hline
\textbf{Bit-cell Area ($\mu$m\textsuperscript{2})} & 2.35 & 4.1 & 4.278 & 2.83 & 4.5 & 2.25 & 2.63 \\ \hline
\textbf{Input width (bits)} & 1-8 & 4/8 & 4/8/12/16 & 1-16 & 4 & 4/8 & 1-64 \\ \hline
\textbf{Weight width (bits)} & 1-8 & 1-8 & 4/8/12/16 & 1-16 & 1-4 & 4/8/12/16 & 1-64 \\ \hline
\textbf{Neural network} & - & LeNet-5 & ResNet-18 & Inception-V4 & ResNet-20 & NA & Inception-V4 \\ \hline
\textbf{Accuracy (\%)} & - & 99.1 & 94.8 & 95.3 & 98.67 & - & 95.6 \\ \hline
\textbf{Throughput (TOPS)} & 0.82* & 2.2 & - & 2.05* & 0.82 & 2.52* & 1.93* \\ \hline
\textbf{\begin{tabular}[c]{@{}l@{}}Energy Efficiency\\ (TOPS/W)\end{tabular}} & 249.1* & 480 & 52.22* & 63* & 273 & 404* & 364* \\ \hline
\end{tabular}}
\end{table*}

The bit-cell is validated across different PVT Corners (TT, FF, SS, SF, FS), temperature ranges (-40, 27, and 125 °C), and VDD varied from 0.6 to 1.2 for both PIM and CAM operations, which were found to be functional. We have reported the performance for TT/27 °C in Fig. 5. Further, we have compared the accumulation with different adder-tree/compressor tree approaches from state-of-the-art designs in terms of delay, power and number of transistors. As shown in Fig. 6, our approach consumes a lower transistor count compared to the conventional design and [22], and achieves significantly faster operation, up to 1.28 $\times$ and 1.45 $\times$ faster than the best state-of-the-art designs [15] and [23], respectively. The design consumes relatively higher power compared to the approximate adder tree work [16], which reflects the accuracy-resource consumption trade-off. 

We developed a parameterised software evaluation model using the $QKeras$ library with Python 3.0 on the Google Colab platform and extracted weights in CSV format for manual mapping to the proposed macro for TinyML acceleration following [10]. The software-evaluation model was equipped with a bit-precision variable error-resilient feature to assess the accuracy drop associated with quantisation and 40\% pruning, reducing additional MAC computations for improved energy efficiency. The hardware performance emulation model draws its references from the following design parameters: parallel multiplications per tile, activation and weight precision, the number of memory banks required for layer execution, and reconfigurable hardware multiplexing. Our evaluation shows 95.6\% accuracy for the Inception v4 model, 91.7\% and 90.34\% for ResNet-18 and VGG-16, respectively, on the CIFAR-10 dataset. This compares to the FP-32 baseline of 93.2\% and 92.64\% for ResNet-18 and VGG-16, using the CIFAR-10 dataset, confirming a Quality of Result (QoR) factor greater than 97.5\%, which outperforms prior approaches [2], [13], [33] by a comparable margin. The detailed macro performance is compared in Table II. Our work shows significantly higher throughput and energy efficiency at the post-layout level compared to [10], [15], [34]. Prior works [13], [35] report pre-layout simulation results, which slightly deviate from the performance translation in terms of throughput and energy efficiency. Our macro exhibits significantly comparable throughput despite incorporating additional CAM operation, with support for various NAF workloads, including Softmax, sigmoid, and tanh, making this macro more useful with LUT [36] for advanced AI computations, where these workloads constitute a growing proportion compared to MAC compute. We mark this exploration as a future scope of this work.

\section{Conclusion \& Future Work}
This work introduced FERMI-ML, a flexible and resource-efficient Memory-In-Situ SRAM macro tailored for TinyML acceleration. The design combines a 9T XNOR-based RX9T bit-cell and a C22T compressor-tree accumulator, achieving a scalable, low-power, and precision-configurable compute fabric for in-memory MAC and CAM operations. Post-layout evaluation confirmed robust functionality across PVT corners, as well as superior energy efficiency and compute density compared to state-of-the-art digital PIM designs. The proposed macro offers high throughput (1.93 TOPS) and energy efficiency (364 TOPS/W), achieving 97.5\% baseline accuracy on benchmark models. Furthermore, the unified PIM-CAM framework extends the application to non-linear activation workloads with LUT-based functions for sigmoid, tanh, and softmax. The extension would focus on multi-bank parallelism and seamless integration into RISC-V-based Edge-AI SoCs as an L3 cache for on-chip learning and inference in next-generation XR platforms.


\end{document}